\documentclass[aps,prb,twocolumn,showpacs,amsmath,amssymb,superscriptaddress,citeautoscript,10pt]{revtex4-2}

\usepackage{graphicx}
\usepackage{siunitx}
\usepackage{hyperref}
\usepackage{bm}
\usepackage{color}
\usepackage[dvipsnames]{xcolor}
\usepackage[version=4]{mhchem}
\usepackage[normalem]{ulem}

\let\originaltextcolor\textcolor

\renewcommand{\textcolor}[2]{%
\ifstrequal{#1}{orange}{\originaltextcolor{black}{#2}}{\originaltextcolor{#1}{#2}}
}

\begin{document}

\title{Uniaxial stress tuning of the anomalous Hall effect in Mn$_3$Ge}

\author{G.~A.~Lombardi}
\affiliation{Max Planck Institute for Chemical Physics of Solids, N\"{o}thnitzer Str.\ 40, 01187 Dresden, Germany}
\affiliation{Brazilian Synchrotron Light Laboratory (LNLS), Brazilian Center for Research in Energy and Materials (CNPEM), Campinas 13083-100, S\~{a}o Paulo, Brazil}
\affiliation{Instituto de F\'{i}sica Gleb Wataghin, Universidade Estadual de Campinas (UNICAMP), Campinas 13083-859, S\~{a}o Paulo, Brazil}

\author{L.~O.~Kutelak}
\affiliation{Brazilian Synchrotron Light Laboratory (LNLS), Brazilian Center for Research in Energy and Materials (CNPEM), Campinas 13083-100, S\~{a}o Paulo, Brazil}
\affiliation{Instituto de F\'{i}sica Gleb Wataghin, Universidade Estadual de Campinas (UNICAMP), Campinas 13083-859, S\~{a}o Paulo, Brazil}

\author{M.~M.~Piva}
\affiliation{Max Planck Institute for Chemical Physics of Solids, N\"{o}thnitzer Str.\ 40, 01187 Dresden, Germany}

\author{V.~E.~S.~Frehse}
\affiliation{Brazilian Synchrotron Light Laboratory (LNLS), Brazilian Center for Research in Energy and Materials (CNPEM), Campinas 13083-100, S\~{a}o Paulo, Brazil}
\affiliation{Instituto de F\'{i}sica Gleb Wataghin, Universidade Estadual de Campinas (UNICAMP), Campinas 13083-859, S\~{a}o Paulo, Brazil}
\affiliation{Center for Electronic Correlations and Magnetism (EKM), Universitätsstraße 1, 86159 Augsburg, Germany}

\author{G.~A.~Calligaris}
\affiliation{Brazilian Synchrotron Light Laboratory (LNLS), Brazilian Center for Research in Energy and Materials (CNPEM), Campinas 13083-100, S\~{a}o Paulo, Brazil}

\author{K.~Manna}
\affiliation{Max Planck Institute for Chemical Physics of Solids, N\"{o}thnitzer Str.\ 40, 01187 Dresden, Germany}
\affiliation{Indian Institute of Technology - Delhi, Hauz Khas, New Delhi 110 016, India}

\author{C.~Felser}
\affiliation{Max Planck Institute for Chemical Physics of Solids, N\"{o}thnitzer Str.\ 40, 01187 Dresden, Germany}

\author{R.~D.~dos Reis}
\email{ricardo.reis@lnls.br}
\affiliation{Brazilian Synchrotron Light Laboratory (LNLS), Brazilian Center for Research in Energy and Materials (CNPEM), Campinas 13083-100, S\~{a}o Paulo, Brazil}

\author{M.~Nicklas}
\email{michael.nicklas@cpfs.mpg.de}
\affiliation{Max Planck Institute for Chemical Physics of Solids, N\"{o}thnitzer Str.\ 40, 01187 Dresden, Germany}

\begin{abstract}

Tunable electronic properties in magnetic materials lead to novel physical phenomena that have the potential to be exploited in the design of new spintronic devices. Here, we report the effect of uniaxial stress on the anomalous Hall effect (AHE) in the hexagonal frustrated antiferromagnetic Heusler compound Mn$_3$Ge. Our x-ray diffraction results show that the $c/a$ ratio varies linearly with strain when stress is applied along the $a$ axis, as well as a significantly higher Young's modulus along the $c$ direction. The linear behavior of the $c/a$ ratio under uniaxial stress mirrors that seen under hydrostatic pressure up to 1.8~GPa, but results in a characteristically different behavior of the AHE. Stress applied along the $a$ axis induces a distortion in the $ab$ plane, smoothing the abrupt jump in the AHE signal at zero magnetic field. In contrast, stress applied along the $c$ axis has little effect, presumably due to the higher Young's modulus. We argue that this is due to pronounced changes in magnetic order.

\end{abstract}

\date{\today}
\maketitle

\textcolor{orange}{\section{INTRODUCTION}}

Spintronics, which exploits the spin degree of freedom in electronic devices, has traditionally focused on ferromagnetic (FM) materials \cite{Novoselov,Chappert_nmat207,Kent_Nat_nano2015,Waldrop_Nature,Enke_natphys2018,Faleev_PhysRevMaterials,Jungwirth_Nature_Phys18,Nayak_2016,Wadley587,Xiao2012,Manna2018}. However, antiferromagnetic (AFM) materials aim to replace ferromagnets in spintronic devices due to their resistance to perturbations, lack of stray fields, and suitability for high-density memory integration \cite{jungwirth2016antiferromagnetic, baltz2018antiferromagnetic, yan2019piezoelectric, magnetochemistry2022, dal2024antiferromagnetic}. In addition, they exhibit ultrafast dynamics and significant magneto-transport effects, making them essential for applications such as nonvolatile memory and magnetic field probes \cite{XIONG2022522}.

The Mn$_3X$ family ($X$ = Ge, Sn, Ga, Ir, Rh, and Pt) hosts several candidates for such applications due to their antiferromagnetic order and chiral nature in transport phenomena \cite{Novoselov,Chappert_nmat207,Kent_Nat_nano2015,Waldrop_Nature,Enke_natphys2018,Faleev_PhysRevMaterials}. Among these, Mn$_3$Ge has a well-established history of exhibiting non-trivial \textcolor{orange}{electronic properties}, driven by its chiral magnetic field generated by canted spins in the kagome lattice \cite{Nayak_2016}. Hydrostatic pressure applied to Mn$_3$Ge can progressively tune the anomalous Hall effect (AHE) signal, even leading to a signal reversion above a critical pressure \cite{dos2020pressure}. \textcolor{orange}{Recently, uniaxial stress has proven to be an effective tool for tuning the electronic and magnetic properties of kagome materials \cite{faria2023sensitivity, aoyama2024piezomagnetic, mojarro2024tuning, lin2024uniaxial}. These advances make uniaxial stress a promising method for precisely controlling the chiral properties and related effects in Mn$_3$Ge for use in novel spintronic devices.}

Mn$_3$Ge is a Heusler compound that can crystallize in hexagonal \textcolor{orange}{($P6_3/mmc$)} and tetragonal \textcolor{orange}{($I4/mmm$)} structures \cite{Zhang_2013}. This study focuses on the hexagonal structure, which exhibits triangular AFM ordering with $T_N$ ranging from 365~K to 400~K \cite{Ohoyama_JPSJP}. In this structure, the unit cell of Mn$_3$Ge consists of two layers of Mn stacked along the $c$ axis. Within each layer, the Mn atoms form a kagome lattice, with Ge positioned at the center of each hexagon.

Recent theoretical and experimental studies have emphasized the potential of strain and stress to modify the electronic and magnetic properties of Mn$_3$Ge. One theoretical work predicted that both in-plane and out-of-plane stress can effectively tune the Hall effect in hexagonal Mn$_3$Ge \textcolor{orange}{by introducing an easy-axis anisotropy within the kagome lattice, which leads to a net strain-dependent moment \cite{dasgupta2022tuning}. Another} proposed that chiral orders form below 200~K and that magnetic moments resist directional changes under rotating magnetic fields in the absence of crystalline defects \cite{chaudhary2022magnetism}. Experimentally, strain of $\approx0.1\%$ was shown to tune the AHE in Mn$_3$Sn \cite{ikhlas2022piezomagnetic}, and epitaxial strain modulated the AHE in both Mn$_3$Sn \cite{wang2019integration} and Mn$_3$Ga \cite{guo2020giant}.

In the present study, we investigate the effects of \textcolor{orange}{in-plane and out-of-plane} uniaxial stress on the AHE in Mn$_3$Ge, taking advantage of the strain sensitive chiral properties of the material. By exploring these effects, we aim to extend the tunability of Mn$_3$Ge for spintronic applications.

\begin{figure*}[tbh]
\begin{center}
 \includegraphics[width=0.9\textwidth]{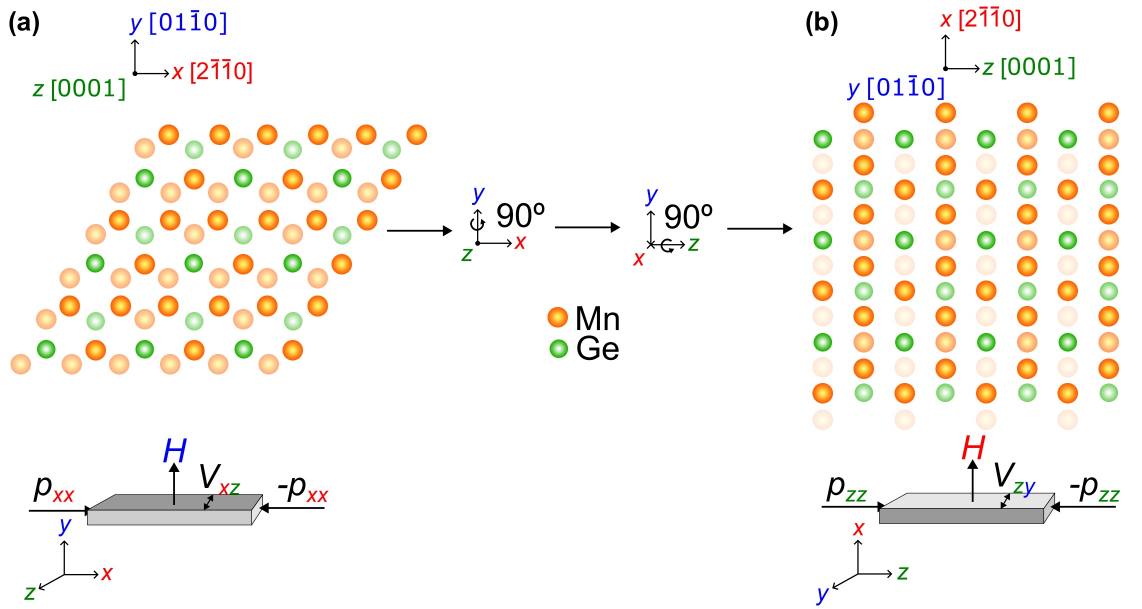}
 \caption{Real space structures of Mn$_3$Ge illustrating two experiments with stress applied along different directions: (a) Stress was applied along the $[2~\overline{1}~\overline{1}~0]$ direction, i.e., along $[1~0~\overline{1}~0]$ ($x$ direction) in real space. The magnetic field was applied perpendicular to the $x$ axis, in the $y$ direction, aligned with the $[0~1~\overline{1}~0]$. Finally, the Hall voltage ($V_{xz}$) was measured along the $[0~0~0~1]$ direction, referred to as the $z$ axis in real space. The second configuration is obtained by performing two successive counterclockwise $90^{\circ}$ rotations on the real lattice. First a rotation around the $y$ axis, followed by a second rotation around the $z$ axis. (b) Stress was then applied along the $z$ axis while the magnetic field was directed along the $x$ axis. Hall voltage ($V_{zy}$) was measured along the $y$ axis.}
 \label{fig:lattice_measurement_scheme}
 \end{center}
\end{figure*}

\textcolor{orange}{\section{METHODS}}

High quality \textcolor{orange}{hexagonal} Mn$_3$Ge single crystals were grown using the Bridgman-Stockbarger technique \cite{Nayak_2016}. \textcolor{orange}{Energy-dispersive x-ray spectroscopy (EDX) measurements revealed a sample homogeneity with atomic percentages of Ge between 23.88-25.66\%}. Single crystal x-ray diffraction (XRD) was performed at the EMA beamline (Sirius) with a 25.514~keV beam focused to a $100\times100~{\rm \mu m^2}$ spot. A Razorbill CS200T strain cell on a 6-axes diffractometer applied uniaxial stress and selected reflections were tracked with a Pilatus 300~K CCD detector \cite{dosReis_2020}. Strain values were determined from the peak shifts with a zero-strain reference corresponding to no voltage applied on the piezo stacks. Electrical transport was measured using a five-point geometry with an LR-700 resistance bridge. Temperature and magnetic field were controlled with a Quantum Design PPMS. \textcolor{orange}{The transverse resistivity was antisymmetrized to isolate the Hall contribution}. Stress was applied using a Razorbill CS100 strain cell, and length change was monitored using the integrated capacitive sensor and an Andeen-Hagerling AH-2500A capacitance bridge. \textcolor{orange}{Two bar-shaped samples were investigated in the electrical transport experiments (see Fig.~\ref{fig:lattice_measurement_scheme} for details). Both samples had the same thickness  $t=170\,\mu\mathrm{m}$ and width $w= 250\,\mu\mathrm{m}$. The free distances between the claws of the strain cell were $350\,\mu\mathrm{m}$ for sample 1 and $890\,\mu\mathrm{m}$ for sample 2. For the Hall experiments, the voltage contacts were positioned on the homogeneously strained part of the sample. Following Hicks \textit{et al.} \cite{hicks2014piezoelectric} we could estimate a strain inhomogeneity across the sample thickness of less than $1\%$.}

\textcolor{orange}{\section{RESULTS}}

\textcolor{orange}{We have studied the effect of uniaxial strain on the crystal structure and the AHE  in Mn$_3$Ge for two characteristically different configurations, applying stress within and perpendicular to the kagome plane. In both cases, the magnetic field was applied parallel to the kagome plane (see Fig.~\ref{fig:lattice_measurement_scheme}). In sample 1}, the stress was applied in the basal plane parallel to the current\textcolor{orange}{, along the $[2~\overline{1}~\overline{1}~0]$ direction. This direction} corresponds to the $a$ axis in real space and is defined as the $x$ direction in the electrical transport experiments. The magnetic field was applied perpendicular to the stress and current directions along $[0~1~\overline{1}~0]$, defined here as the $y$ direction ($b^*$ axis). \textcolor{orange}{In sample 2}, the stress and current are applied perpendicular to the basal plane along the $[0~0~0~1]$ direction ($c$ axis), defined here as the $z$ direction. In this case, the magnetic field is applied along the $x$ direction. We observe a clear difference in the behavior of the AHE in these two different configurations. Before discussing the AHE data obtained under uniaxial strain in detail, we will first examine the structural changes with applied stress.

\begin{figure}[tbh!]
\begin{center}
 \includegraphics[clip,width=0.85\columnwidth]{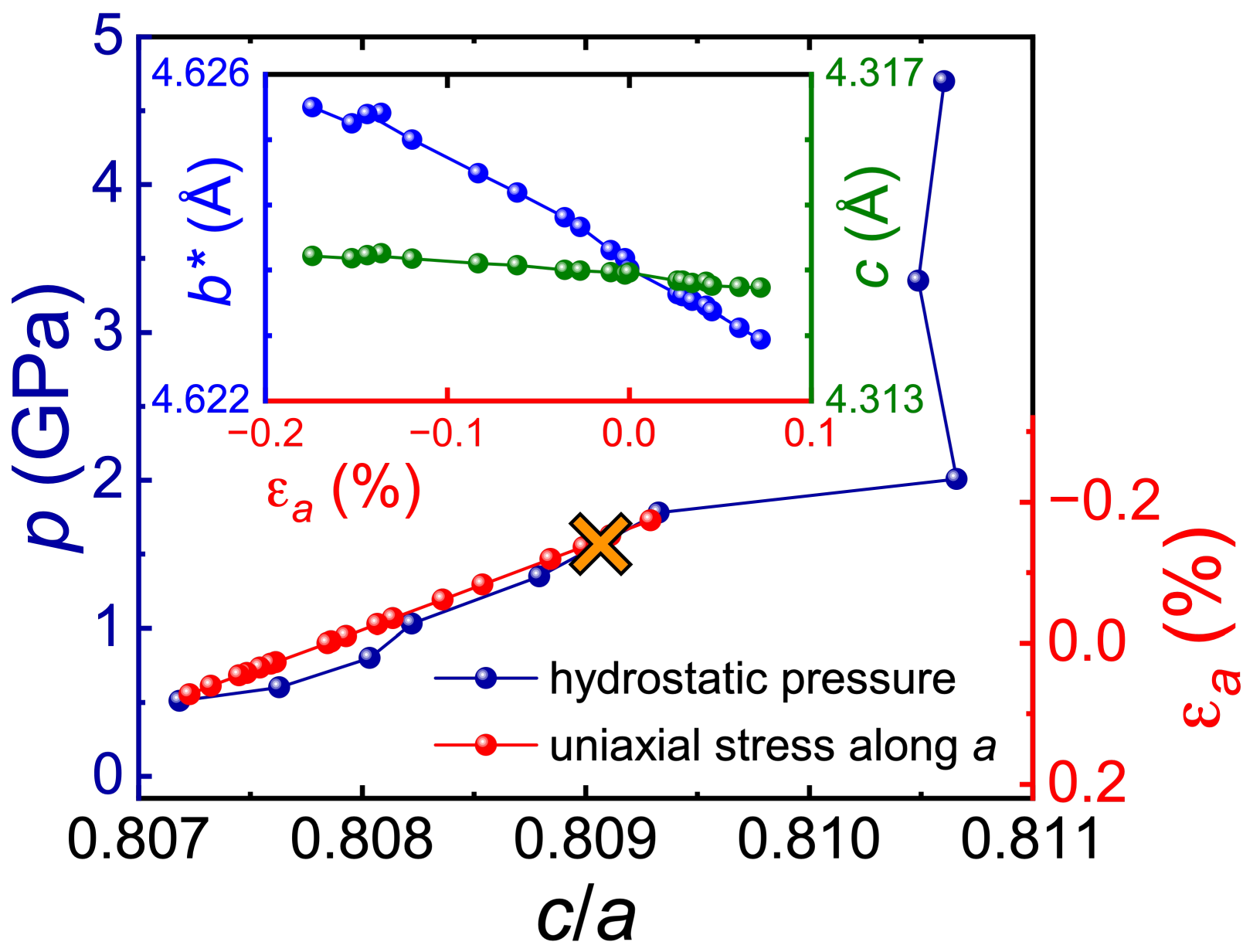}
 \caption{XRD results under applied stress along the $a$ direction are compared with hydrostatic pressure data extracted from \cite{sukhanov2018gradual}, both at 300~K. The right axis of the plot (red spheres) shows the relative strain along $a$, $\varepsilon_a$, as a function of the lattice parameter ratio $c/a$, where zero strain corresponds to the nominal experimental value. The left axis (blue spheres) represents the applied hydrostatic pressure as a function of $c/a$. \textcolor{orange}{The "X" marks the point at which the AHE is suppressed under hydrostatic pressure.} The inset shows the variation of the interplanar planes $d_{01\overline{1}0}$, corresponding to $b^*$, and $d_{0001}$, corresponding to the lattice parameter $c$, as a function of strain under stress applied along the $a$ direction. The step size is the same for the left and right axes.}
 \label{fig:xrd}
 \end{center}
\end{figure}

XRD experiments under uniaxial stress applied along the $a$ direction allow us to compare our AHE results with previous results under hydrostatic pressure based on the crystal structure \cite{dos2020pressure}. The strain induced in the sample by the stress applied along the $a$ direction $\varepsilon_a$ was determined by measuring the shifts of the diffraction peaks corresponding to the $(\overline{12}~6~6~0)$, $(0~\overline{12}~12~0)$, and $(0~0~0~\overline{8})$ reflections. These reflections correspond to crystallographic directions collinear to the $x$, $y$, and $z$ directions, respectively, used in the electrical transport experiments (see Fig.~\ref{fig:lattice_measurement_scheme}) and were accessible due to the experimental geometry at the beamline. A maximum compressive strain of about $-0.17\%$ was achieved with no structural phase transition observed. 

The inset of Fig.~\ref{fig:xrd} shows the variation of $b^*$ and $c$, corresponding to the interplanar distances $d_{01\overline{1}0}$ and $d_{0001}$, respectively, as a function of $\varepsilon_a$. The data are consistent with the significant difference in elastic moduli, with the $c$ axis having  a larger Young's modulus \cite{theuss2022strong}. While stress in the $a$ direction induces a large change in the lattice parameters $a$ and $b^*$, the variation in $c$ is an order of magnitude smaller. For all three directions the change is linear in $\varepsilon_a$ with a slope of $0.05~{\rm\AA}/\%$, $-0.01~{\rm \AA}/\%$, and $-0.002~{\rm \AA}/\%$ for $a$, $b^*$, and $c$, respectively. This difference suggests that a much larger stress must be applied along the $c$ axis ($z$ direction) to induce an observable effect in the AHE. 

A comparison of the $c/a$ ratio under hydrostatic pressure (taken from \cite{sukhanov2018gradual}) and uniaxial stress applied along the $a$ direction is shown in Fig.~\ref{fig:xrd}. Under hydrostatic pressure, the ratio $c/a$ varies linearly with pressure up to about 1.8~GPa. Beyond this pressure, $c/a$ remains constant due to a magnetostriction effect induced by the out-of-plane canting of the magnetic structure \cite{sukhanov2018gradual,dos2020pressure}. When uniaxial stress is applied along the $a$ direction, the resulting strain, which ranges from $-0.17\%$ compressive to $0.07\%$ tensile, leads to a $c/a$ ratio that is linear in $\varepsilon_a$. The linear dependence is consistent with the linear range observed under hydrostatic pressure. This allows us to directly compare the AHE data obtained under hydrostatic pressure and uniaxial stress based on the changes in lattice structure.

\begin{figure}[tb!]
\begin{center}
 \includegraphics[clip,width=0.9\columnwidth]{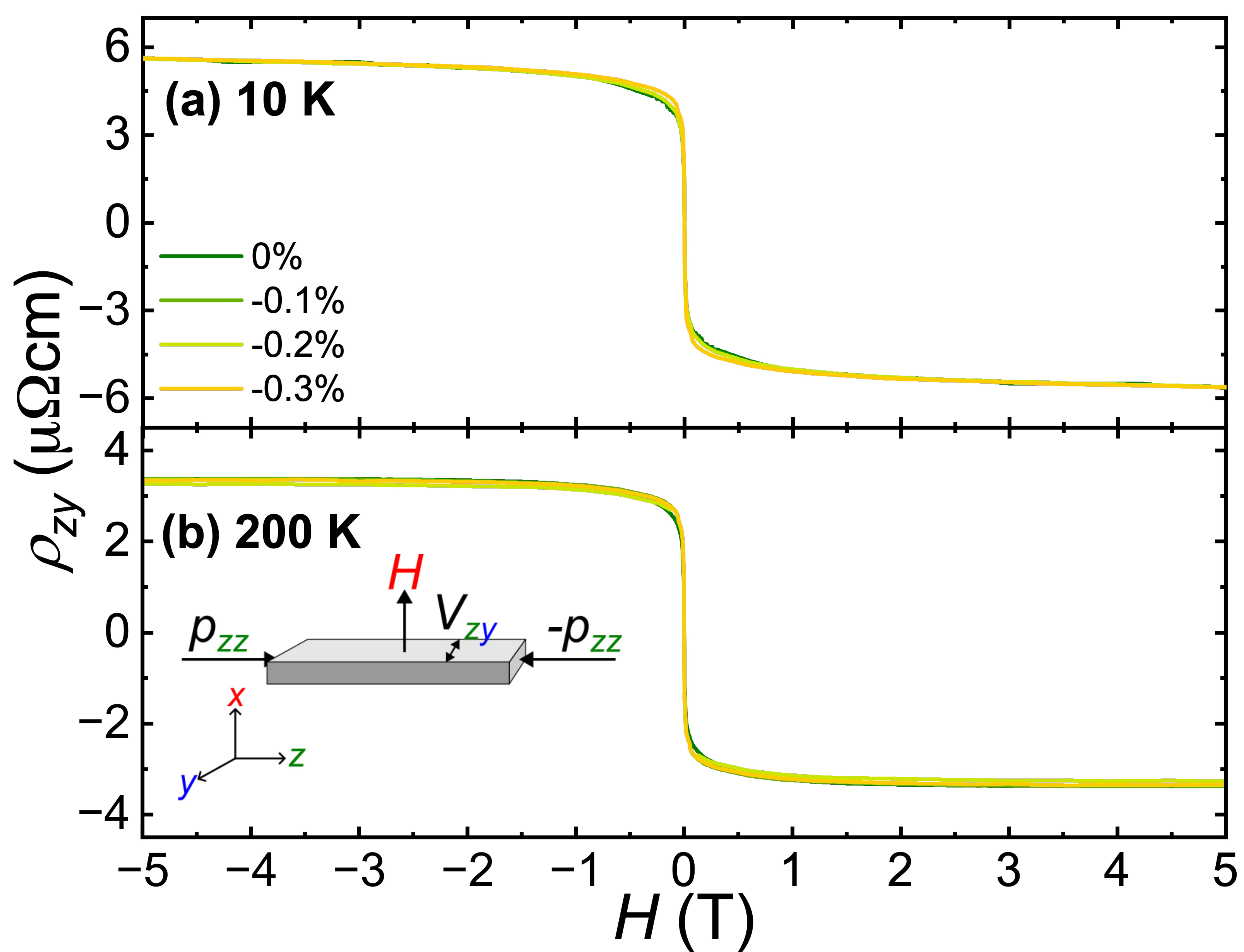}
 \caption{Hall resistivity $\rho_{zy}$ of Mn$_3$Ge under compressive stress applied along the $z$ axis at (a) 10~K and (b) 200~K. The field was applied along the $y$ axis \textcolor{orange}{and the data was recorded with an increasing magnetic field}. The zero strain curves were measured separately outside of the strain cell.
 }\label{fig:hall_strain_direcao_c}
 \end{center}
\end{figure}

\begin{figure}[tb!]
\begin{center}
 \includegraphics[clip,width=0.9\columnwidth]{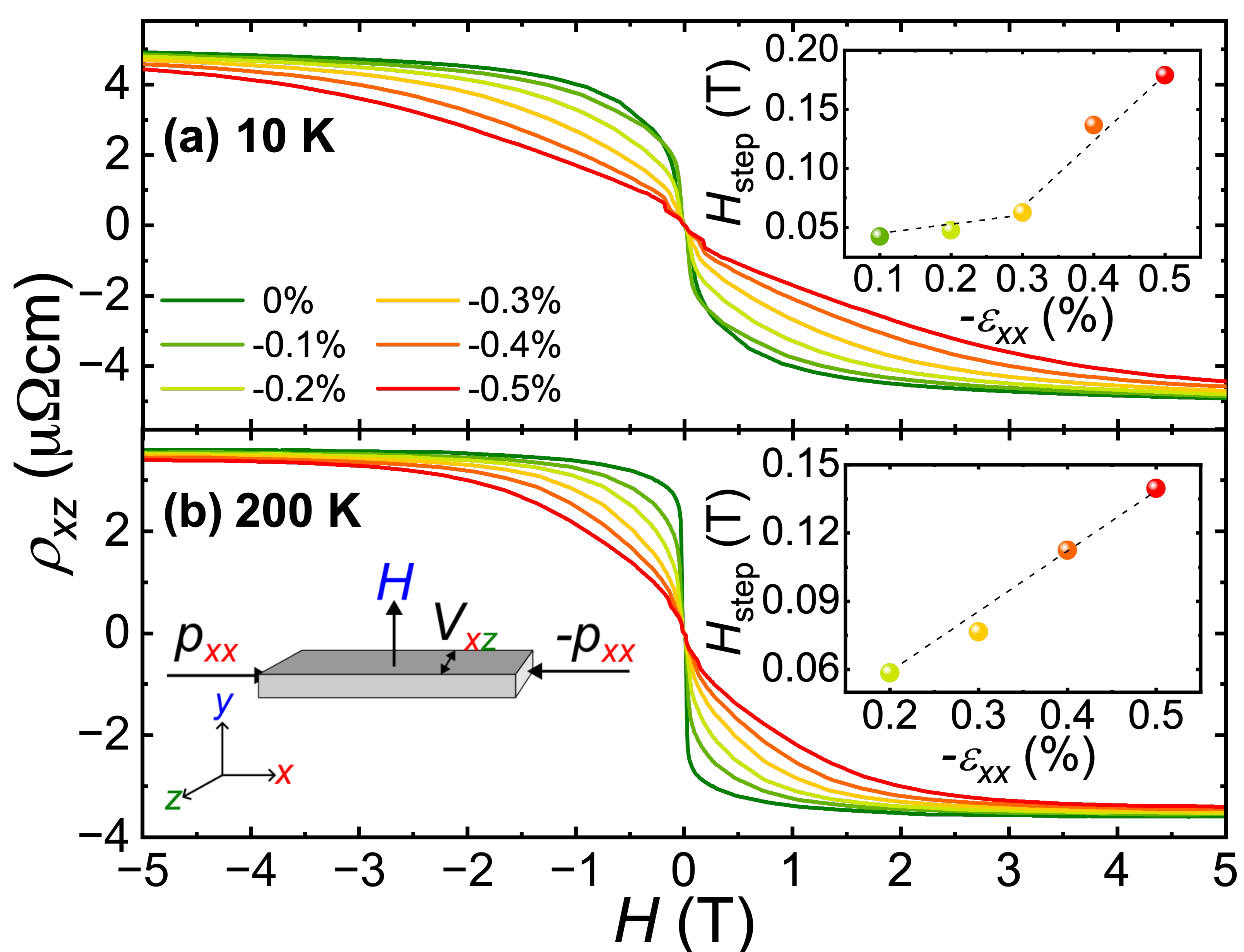}
 \caption{Hall resistivity $\rho_{xz}$ of Mn$_3$Ge under compressive stress applied along the $x$ axis at (a) 10~K and (b) 200~K. The magnetic field was applied along the $y$ direction \textcolor{orange}{and the data was recorded with an increasing magnetic field}. The zero strain curve at 10~K was measured separately outside of the strain cell. The insets \textcolor{orange}{display the positions of the step-like features observed at low magnetic fields as a function of compressive strain. The dotted lines serve as a guide to the eye}.
 }\label{fig:hall_strain_plano_ab}
 \end{center}
\end{figure}

We now turn to the strain evolution of the AHE. The antisymmetrized Hall resistivity $\rho_{zy}$, \textcolor{orange}{plotted for increasing field}, under stress applied \textcolor{orange}{perpendicular to the kagome plane} along the $z$ axis at 10~K and 200~K is presented in Fig.~\ref{fig:hall_strain_direcao_c}. $\rho_{zy}$ shows a significant anomalous Hall effect with a step-like change from positive values at negative fields to negative values at positive fields. \textcolor{orange}{The ordinary Hall effect is small. It is negligible compared to the total Hall signal (see the Appendix).} Applying stress does not lead to significant changes in the AHE for compressive strains up to $-0.3\%$. At 200~K we find saturation of the Hall resistivity at $|\rho_{zy}^{\rm sat}|\approx 5.6~{\rm \mu \Omega cm}$ for high magnetic fields, while at 10~K $\rho_{zy}$ tends to saturate but still shows a finite slope at $\pm 5$~T. At $\pm5$~T we find $|\rho_{zy}^{\rm sat}|\approx 3.4~{\rm \mu \Omega cm}$. The finding that the AHE is insensitive to stress applied along the $z$ axis in Mn$_3$Ge is in contrast to hydrostatic pressure experiments where the Hall resistivity is probed in the same measurement configuration (see Fig.~\ref{fig:hall_strain_direcao_c}), but is consistent with results on the closely related material Mn$_3$Sn \cite{ikhlas2022piezomagnetic}, which is also insensitive to out-of-plane stress.

In contrast to out-of-plane stress, in-plane stress has a strong effect on the AHE. Fig.~\ref{fig:hall_strain_plano_ab} shows the antisymmetrized $\rho_{xz}$ in applied magnetic field between -5~T and 5~T under different compressive stresses applied along the $x$ direction for 10~K and 200~K. At zero strain $\rho_{xz}(H)$ shows a similar behavior as described above for $\rho_{zy}(H)$. $\rho_{xz}$ is positive for negative fields and shows a step-like feature at zero field with the signal reversing at positive fields. However, especially at low temperatures, the step is not as sharp as in $\rho_{zy}(H)$. In addition, we find a clear strain dependence of the AHE for both temperatures. At 10~K, the Hall resistivity saturates around $|\rho_{xz}^{\rm sat}|\approx 4.5~\mu \Omega {\rm cm}$ above $\pm2$~T. Increasing compressive strain shifts the field above which $\rho_{xz}(H)$ saturates to higher values, while simultaneously broadening the step in the AHE signal. At maximum strain we find an almost linear field dependence of $\rho_{xz}$ around zero field. $|\rho_{xz}^{\rm sat}|$ drops from about 4.9~$\mu\Omega {\rm cm}$ at zero strain to 4.4~$\mu\Omega {\rm cm}$ at $-0.5\%$ compressive strain at 5~T. At 1~T, the value drops from $|\rho_{xz}^{\rm sat}|\approx 4~\mu\Omega {\rm cm}$ to 1.7~$\mu\Omega {\rm cm}$ in the same strain range. This behavior seems to be in contrast to that observed under hydrostatic pressure, where increasing pressure leads to an inversion of the AHE signal around 1.53~GPa, at which pressure the AHE signal disappears \cite{dos2020pressure}.

At 10~K, a prominent kink appears in the Hall resistivity curve around \textcolor{orange}{$\pm0.04$~T}. This kink shifts monotonically to higher fields with increasing strain. \textcolor{orange}{The field position of these kinks ($H_{step}$) as a function of compressive strain is shown in the inset of Fig.~\ref{fig:hall_strain_plano_ab}(a)}. A similar feature can be observed at $\SI{200}{\kelvin}$ starting at $-0.2\%$ strain at about $\pm0.06$~T [inset Fig.~\ref{fig:hall_strain_plano_ab}(b)]. Although not as pronounced as at 10~K, there is a linear shift to higher fields with increasing strain. The possible origin of this feature is discussed below.

\begin{figure}[tb!]
\begin{center}
 \includegraphics[clip,width=0.9\columnwidth]{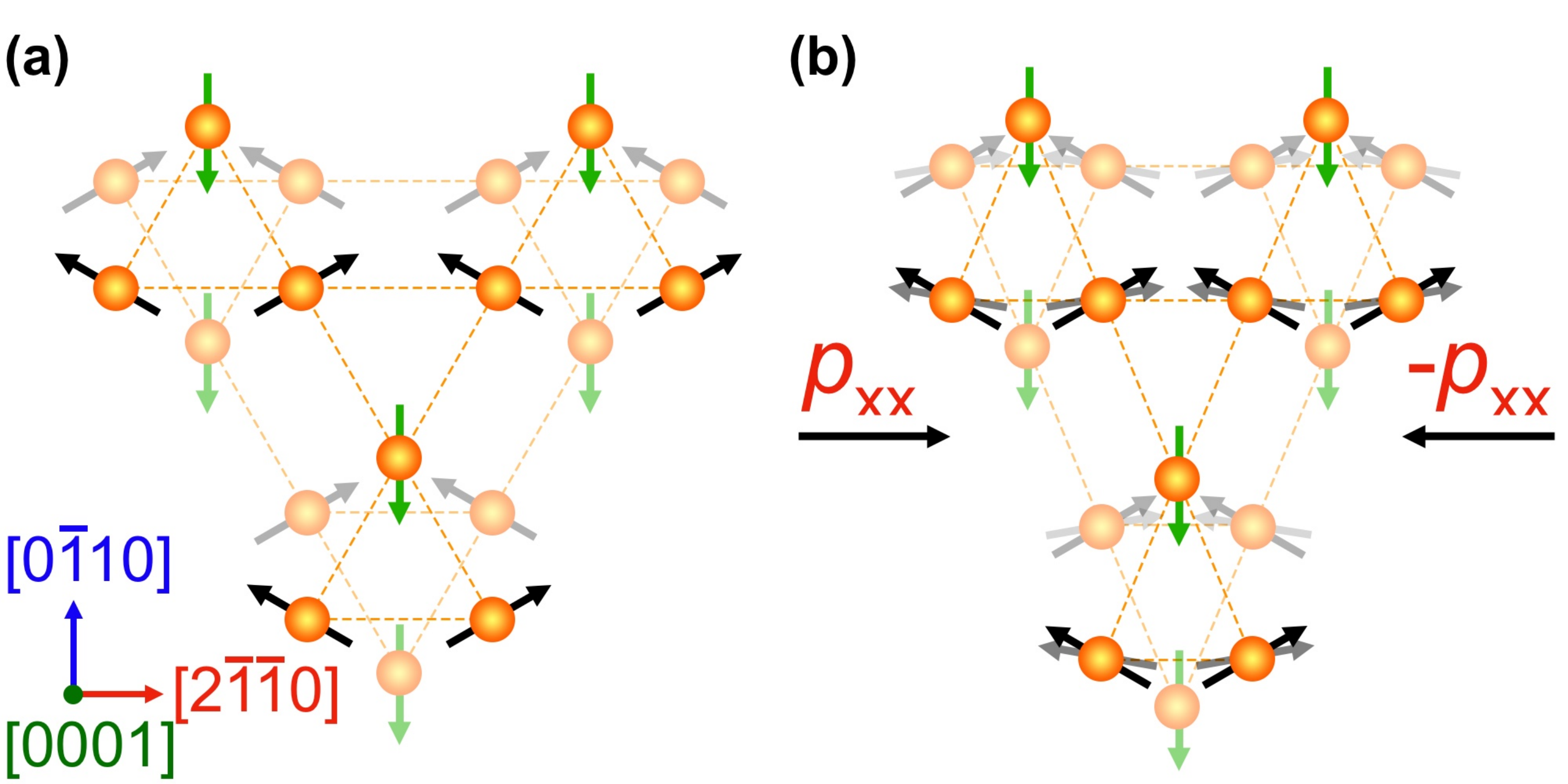}
 \caption{\textcolor{orange}{(a) Magnetic structure of Mn$_3$Ge for a magnetic field applied along the $[0~1~\overline{1}~0]$ direction. The black and green arrows represent the anti-chiral antiferromagnetic structure. When the field direction is reversed, the spins flip $180^{\circ}$ \cite{kiyohara2016giant}. The green arrows correspond to the moments aligned with the local easy axis \cite{dasgupta2020theory}. (b) Stress induces an additional easy axis along the stress application that induces the spins to rotate towards such axis.}}
 \label{fig:magn_struc}
 \end{center}
\end{figure}

\textcolor{orange}{\section{DISCUSSION}}

Mn$_3$Ge has a ground state magnetic structure where the local moments are arranged in an AFM antichiral $120^\circ$ triangular configuration \textcolor{orange}{[see Fig.~\ref{fig:magn_struc}(a)]} driven by Heisenberg exchange interactions between nearest neighbors \cite{Nagamya1979, NAGAMIYA1982}. The chirality of this arrangement is determined by a Dzyaloshinskii-Moriya (DM) interaction \cite{dzyaloshinsky1958thermodynamic, moriya1960anisotropic}. This interaction constrains the spins to lie within the kagome plane, with no discernible canting along the $c$ axis. The Ge site introduces a weak local easy-axis anisotropy that is significantly smaller than both the exchange and DM interactions. This anisotropy perturbs the $120^\circ$ spin configuration in the kagome plane, resulting in a small dipolar FM moment of about $10^{-3}\mu_B$ per Mn atom \cite{dasgupta2022tuning}. \textcolor{orange}{In our samples, this residual moment induces a small hysteresis in $\rho_{xz}$, with a coercive field of 200~Oe at 200~K. This result is in excellent agreement with that reported in Ref.~\cite{kiyohara2016giant}. The hysteresis observed in $\rho_{zy}$ is much smaller, with a coercive field of 70~Oe at 200~K, indicating a negligible out-of-plane component of the magnetic moment. However, the small FM in-plane moment has been shown not to play a role in the origin of the AHE \cite{Nayak_2016}.} Previous studies of the angular dependence of the AHE have shown that this residual magnetic moment\textcolor{orange}{, as well as the applied field, solely changes the chirality of the spin structure} \cite{NAGAMIYA1982,Nayak_2016}.

A secondary local easy axis points from the Mn sites toward the center of each triangle \cite{dasgupta2020theory}. Due to the dominant exchange interaction $J$, neighboring spins are locked into a specific pattern, effectively fixing the orientation of the three-spin arrangement in each triangle. Consequently, the order parameter of the system can be described as a vector representing this locked spin configuration. This local symmetry corresponds to the $D_{3h}$ group. Its irreducible representation gives rise to six magnetic modes: three in-plane modes ($\alpha_0$, $\alpha_x$, $\alpha_y$) and three out-of-plane modes ($\beta_0$, $\beta_x$, $\beta_y$) \cite{dasgupta2020theory, dasgupta2022tuning}.

The magnetic ground state of Mn$_3$Ge at ambient pressure corresponds to six of the twelve possible minima of the in-plane $\alpha_0$ mode, which corresponds to uniform rotations of the three spins within the sublattice and does not result in net magnetization per sublattice \cite{dasgupta2020theory,dasgupta2022tuning}. These sixfold Mn$_3$Ge ground states are those in which one of the spins points toward the local easy axis, i.e., toward the center of the triangles of the kagome lattice \cite{tomiyoshi1983triangular,sukhanov2018gradual,soh2020ground}. Application of magnetic field reduces the sixfold degeneracy of the $\alpha_0$ mode to a twofold degeneracy, depending on the field direction \cite{kiyohara2016giant}. This reduction accounts for the step-like signal inversion of the AHE at zero magnetic field observed in our data at ambient pressure (see Figs.~\ref{fig:hall_strain_direcao_c} and \ref{fig:hall_strain_plano_ab}).

When hydrostatic pressure is applied, the $c/a$ ratio increases linearly up to about 1.8~GPa (see Fig.~\ref{fig:xrd}) \cite{sukhanov2018gradual}. In this regime, the AHE signal is continuously suppressed until it disappears at about 1.5~GPa and the AHE signal reverses and starts to increase again \cite{dos2020pressure}. This behavior of the AHE is attributed to a change in the magnetic structure of Mn$_3$Ge, where the Mn spins rotate uniformly out of the plane and acquire a component along the $c$ direction, i.e. hydrostatic pressure favors the $\beta_0$ mode \cite{sukhanov2018gradual, dasgupta2022tuning},  modifying the non-zero Berry curvature.

However, applying uniaxial stress along the $a$ axis does not suppress the AHE signal, although the $c/a$ ratio exhibits a linear dependence on uniaxial strain $\varepsilon_a$ and, \textcolor{orange}{as shown in Fig. \ref{fig:xrd}}, this $c/a$ ratio can be mapped directly onto the data obtained under hydrostatic pressure, indicating similar effects on the crystal structure. Nevertheless, the different responses of the AHE signal to hydrostatic pressure and uniaxial stress suggest that the resulting changes in magnetic structure are fundamentally different, even though the crystal structure is affected in a similar way.

Uniaxial stress introduces an additional easy-axis anisotropy that modifies the magnetic ground state by changing the Heisenberg exchange parameter due to lattice displacements. In-plane strain distorts the $120^\circ$ spin arrangement and generates a strain-dependent net in-plane moment. The intrinsic part of the Hall conductivity tensor is coupled to the Berry curvature via the Hall vector, which acts as a fictitious magnetic field in momentum space \cite{nagaosa2010anomalous}. As shown by Dasgupta \cite{dasgupta2022tuning}, the $\alpha_0$ magnetic mode is linearly coupled to the Hall vector, allowing manipulation of the $120^\circ$ spin configuration to influence the Hall effect.

In contrast to hydrostatic pressure, which forces the spins out of the basal plane resulting in the $\beta_0$ mode \cite{sukhanov2018gradual}, uniaxial stress is expected to modify the anisotropy energy scale and tune the AHE by rotating the spins only in-plane without inducing tilting along the $z$ direction \cite{dasgupta2022tuning}. \textcolor{orange}{An illustration of this effect is shown in Fig.~\ref{fig:magn_struc}(b)}. Moreover, the steps observed in the AHE at low magnetic fields, shown in the insets of Fig.~\ref{fig:hall_strain_plano_ab}, can be explained by the degeneracy breaking of the $\alpha_0$ mode, which depends on the anisotropy energy scale \cite{ikhlas2022piezomagnetic}. As the anisotropy increases with strain, this degeneracy breaking field shifts almost linearly to higher magnetic fields.

The results obtained on Mn$_3$Ge can be compared to those of the closely related compound Mn$_3$Sn, which exhibits a similar strain dependent AHE \cite{ikhlas2022piezomagnetic}. This compound has the same crystal structure as Mn$_3$Ge, but is proposed to have a different magnetic ground state. While the proposed degenerate magnetic ground state of Mn$_3$Ge corresponds to six of the twelve minima of the $\alpha_0$ mode mentioned above, that of Mn$_3$Sn corresponds to the remaining six configurations \cite{dasgupta2020theory,dasgupta2022tuning}. In Mn$_3$Sn the local easy axis does not point to the center of the triangular sublattice as in Mn$_3$Ge, but is aligned along the $a$ direction \cite{dasgupta2020theory}.

AHE data taken on Mn$_3$Sn at room temperature in the same geometry as used here [see Fig.~\ref{fig:lattice_measurement_scheme}(a)] also revealed a strong dependence of the AHE on strain. A signal inversion of the AHE was observed between strains of $-0.062\%$ and $-0.21\%$. In contrast, theoretical predictions from Ref.~\cite{dasgupta2022tuning} suggest that Mn$_3$Ge should exhibit a similar signal inversion at about half of this strain range. However, direct comparison of strain values is not straightforward due to intrinsic material differences. It is known that hexagonal Mn$_3$Ge is only stable with excess Mn randomly occupying the Ge site \cite{yamada1988magnetic, kiyohara2016giant}. This can lead to variations in electronic structure and magnetic anisotropy that affect how strain modifies the band structure and associated Berry curvature. Experimental uncertainties in cell calibration and efficiency of stress transfer to the sample may also be a factor. This needs to be considered when interpreting strain-dependent phenomena in different materials, even within the same structural family. Although the theoretical prediction of Ref.~\cite{dasgupta2022tuning} suggests that stress applied along the $c$ axis should have a similar effect as in-plane stress, the elastic moduli of both compounds differ significantly between in-plane and out-of-plane directions. Mn$_3$Ge has a $c$ axis elastic modulus that is 2.3 times larger than the in-plane modulus one, with a nominal Poisson's ratio $\nu_{ca}$ of 0.041 \cite{theuss2022strong}. \textcolor{orange}{Our measured value, $\nu_{ca} = 0.035 (1)$, is in reasonable agreement with the theoretical value.}

\textcolor{orange}{\section{CONCLUSION}}

In conclusion, our study of the effect of uniaxial stress on the crystal structure and AHE in Mn$_3$Ge reveals a complex interplay between strain, magnetic structure, and electronic properties. We find that uniaxial stress applied along the $a$ axis significantly modifies the AHE, while stress along the $c$ axis has a negligible effect. This difference is attributed to the different effects of in-plane and out-of-plane stress on the magnetic structure, with in-plane stress distorting the $120^\circ$ spin arrangement and generating a strain-dependent net in-plane moment. Our results contrast with those obtained under hydrostatic pressure, where the AHE is reversed. The strain-dependent AHE in Mn$_3$Ge is consistent with theoretical predictions and observations in Mn$_3$Sn. Our results highlight the importance of considering the specific effects of strain on the magnetic structure and electronic transport properties, which are crucial for the design of advanced spintronic devices with tailored properties. The ability to precisely control the AHE through uniaxial strain could enable the development of high-performance magnetic sensors and memory devices with enhanced sensitivity and stability, highlighting the potential of strain engineering in spintronics.\\

\section*{ACKNOWLEDGMENTS}
This project has received funding from the European Union’s Horizon 2020 research and innovation program under Marie Skłodowska-Curie Grant No. 101019024. GAL, RDdR, LOK, and VESF acknowledge financial support from the Brazilian agencies CNPq, CAPES and FAPESP (Grants 140632/2018-2, 2018/00823-0, 2018/19497-6, 2021/02314-9, 2022/05447-2 and 2020/11399-5) and from the Max Planck Society under the auspices of the Max Planck Partner Group R.\ D.\ dos Reis of the MPI for Chemical Physics of Solids, Dresden, Germany. This research used facilities of the Brazilian Synchrotron Light Laboratory (LNLS), part of the Brazilian Center for Research in Energy and Materials (CNPEM), a private non-profit organization under the supervision of the Brazilian Ministry for Science, Technology, and Innovations (MCTI). The staff of EMA and LCTE are acknowledged for their assistance during the 2020227 experiment. CF was financially supported by the Deutsche Forschungsgemeinschaft under SFB 1143 (Project No. 247310070) and the Würzburg-Dresden Cluster of Excellence on Complexity and Topology in Quantum Matter—ct.qmat (EXC 2147, Project No. 39085490).

\section*{DATA AVAILABILITY}

The data that support the findings of this article are openly available \cite{3.UTH7RG_2025}.

\textcolor{orange}{\section*{APPENDIX}}

\begin{figure}[tb!]
\begin{center}
 \includegraphics[clip,width=0.9\columnwidth]{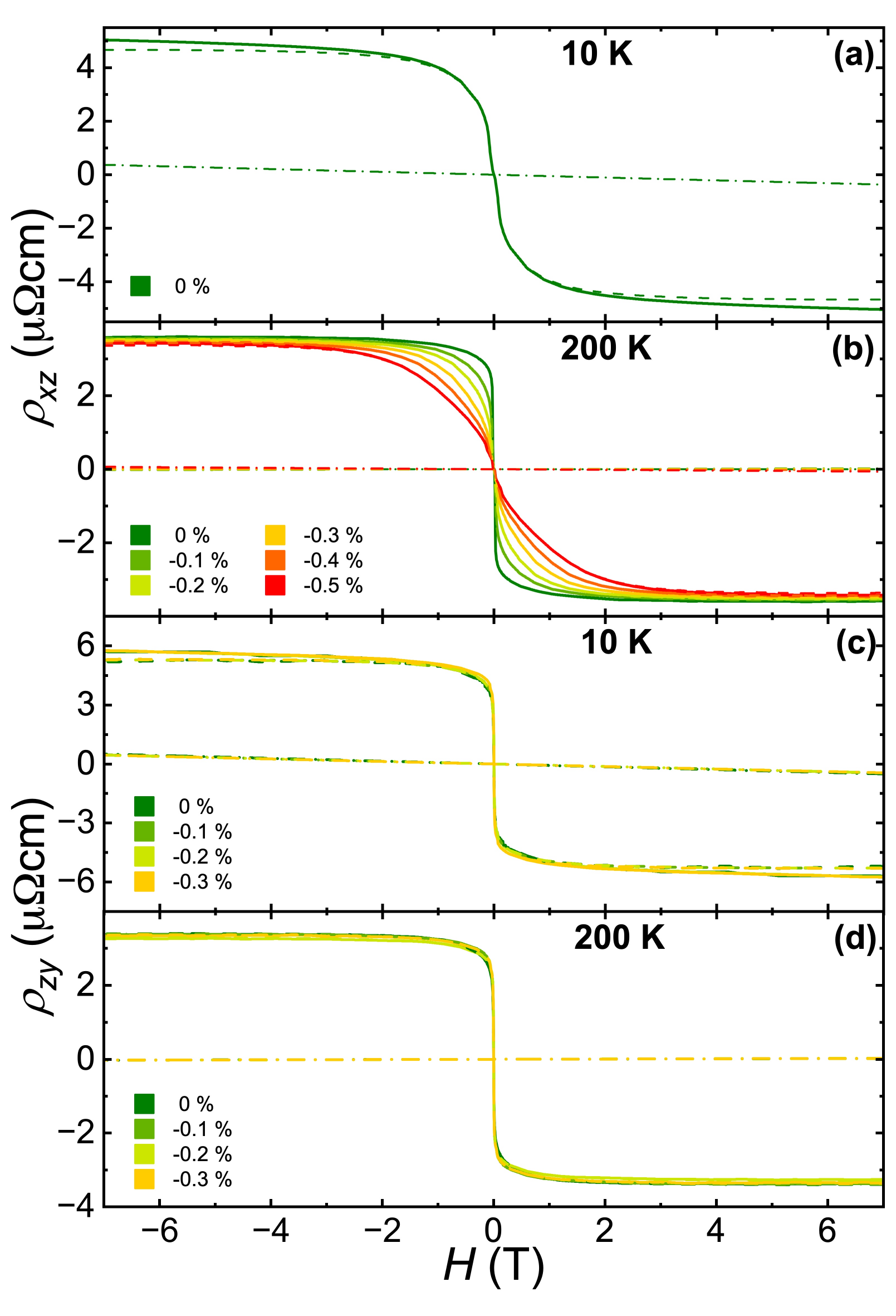}
 \caption{\textcolor{orange}{Anomalous Hall resistivity (dashed lines) derived by subtracting the ordinary Hall contribution (dashed-dotted lines) from the total Hall signal (solid lines). Panels (a) and (b) correspond to measurements with current and stress applied along the $x$ axis and magnetic field along the $y$ axis. Panels (c) and (d) show data with current and stress applied along the $z$ axis and a magnetic field along the $x$ axis.}}
 \label{fig:append1}
 \end{center}
\end{figure}

\textcolor{orange}{The contribution of the ordinary Hall effect to the total Hall effect is small in Mn$_3$Ge. To separate the contributions of the ordinary and anomalous Hall effects we performed a linear fit in the saturation region of the Hall signal at high magnetic fields, as described in Ref.~\cite{rai2024weyl}. We applied the linear fit to the field range between 6~T and 9~T and subtracted the resulting ordinary Hall effect contribution from the total Hall signal. Fig.~\ref{fig:append1} shows the total, anomalous, and ordinary Hall contributions for the two  investigated components, $\rho_{xz}$ and $\rho_{zy}$, at 10~K and 200~K. Note that we lack data above 5~T for $\rho_{xz}$ at 10~K, except at zero strain, which prevents an exact estimation of the ordinary Hall component. }

\newpage

%
\end{document}